\documentclass[a4paper,10pt,twoside]{cpc-hepnp}

\usepackage{multicol}
\usepackage{graphicx}
\usepackage{booktabs}
\usepackage{amssymb,bm,mathrsfs,bbm,amscd}
\usepackage[tbtags]{amsmath}
\usepackage{lastpage}

\begin{document}

\fancyhead[c]{\small Submitted to Chinese Physics C} \fancyfoot[C]{\small 010201-\thepage}

\footnotetext[0]{Received 14 March 2009}

\title{The study of the aging behavior on large area MCP-PMT\thanks{Supported by NSFC(10775181) and Strategic Priority Research
Program of the Chinese Academy of Sciences (XDA10010200 \&
XDA10010400)}}

\author{WANG Wen-Wen$^{a,c}$ \quad QIAN
Sen$^{b,c;b)}$\email{qians@ihep.ac.cn} \quad QI Ming$^{a}$ \quad XIA
Jing-Kai$^{b,c,d}$ \quad CHENG Ya-Ping$^{b,c,d}$  \\\quad NING
Zhe$^{b,c}$ \quad LUO Feng-Jiao$^{b,c,d}$ \quad HENG Yue-Kun$^{b,c}$
\quad LIU Shu-Lin$^{b,c}$ \quad SI Shu-Guang$^{e}$ \\ \quad Sun
Jian-Ning$^{e}$ \quad LI Dong$^{e}$ \quad WANG Xing-Chao$^{e}$ \quad
HUANG Guo-Rui$^{e}$ \quad TIAN Jing-Shou$^{f}$ \\\quad WEI
Yong-Lin$^{f}$ \quad LIU Hu-Lin$^{f}$ \quad LI Wei-Hua$^{f}$
\quad WANG Xing$^{f}$ \quad XIN Li-Wei$^{f}$ \\
 }

\maketitle

\address{%
$^a$ Nanjing Unversity, Nanjing 210093, China\\
$^b$ State Key Laboratory of Particle Detection and Electronics, Beijing 100049, China\\
$^c$ Institute of High Energy Physics, Chinese Academy of Sciences, Beijing 100049, China\\
$^d$ University of Chinese Academy of Sciences, Beijing 100049,China\\
$^e$ Nanjing, North Night Vision Tech. Ltd., NanJing 211106, China\\
$^f$ Xi¡¯an Institute of Optics and Precision Mechanics, Chinese Academy of Sciences, Xi¡¯an 710068,China\\
}

\begin{abstract}
Aging experiments of a novel type of large area MCP-PMT made by JUNO collaboration were conducted. In these aging experiments, the multi-photoelectron spectrum and single photoelectron spectrum were measured daily, as well as the MCP resistance of the second PMT before and after the experiment. Two PMTs were aged successively for cross check. The first PMT was aged for 52 days, while the other one was aged for 84 days. In order to study the mechanism of the aging process, the high voltage on the second PMT was increased to accelerate its aging process when the cumulative output of charge from its anode was about 4 C. From our study, it can be known that large area MCP-PMT aging had a strong relationship with the related MCPs. In accordance with the PMT aging curve, a PMT aging model was setup and a general aging formula was given.
\end{abstract}

\begin{keyword}
Aging, MCP-PMT, large area PMT
\end{keyword}

\begin{pacs}
29.30.-h, 29.40.Mc, 85.60.Ha
\end{pacs}

\footnotetext[0]{\hspace*{-3mm}\raisebox{0.3ex}{$\scriptstyle\copyright$}2013
Chinese Physical Society and the Institute of High Energy Physics
of the Chinese Academy of Sciences and the Institute
of Modern Physics of the Chinese Academy of Sciences and IOP Publishing Ltd}%

\begin{multicols}{2}

\section{Introduction}

The Daya Bay Reactor Neutrino Experiment (Daya Bay) reported that it had measured a non-vanishing value for the neutrino mixing angle¦È13 with a significance of 5.2 standard deviations\cite{lab1,lab2}. Jiangmen Underground Neutrino Observatory (JUNO) is planned to be set up to study the properties of the three generations of neutrinos and the relationship among them \cite{lab3,lab4}. The experiment is supposed to be finished in 2020\cite{lab5}. As an updated experiment based on the Daya Bay Reactor Neutrino Experiment, JUNO has a comparatively higher precision than the former\cite{lab6}. PMTs are the core detectors of JUNO and will be operated continually under high voltage during the entire data taking period. The PMT performances such Gain, Quantum Efficiency (QE), linearity, Collective Efficiency (CE) and energy resolution of a single photo electron spectrum, will change when running for a long time.

Some studies show that aging of PMTs is related to the cumulative output charges of the anode of PMTs \cite{lab7}, thus, the PMT aging has a relationship with the number of input photons and gains. In the experiment conditions of JUNO, the number of input photons is a constant. If other external conditions are unchanged, the higher the gain is, and the higher the speed of aging will be. According to the experiences gained from the Daya Bay experiment, PMTs are expected to operate at a continuous and stable gain of 1.0$\times10^7$.

There are two candidates of PMTs for JUNO \cite{lab8}. One is a 20 inch traditional dynode PMT, the other is a 20 inch MCP-PMT \cite{lab9}.  The multiplying component of the former PMT is based on the dynode, and the latter is based on the MCP. Since the multiplying principle is different, and the aging mechanism should also be different. There are many studies on the aging of dynode PMTs. Sebastiano Aiello \cite{lab7} studied the aging mechanism of this type of PMTs, and claimed that the PMT aging mainly depends on the extent of the last dynode aging. An aging model to analyze dynode PMT aging had also been done by Sebastiano.

The MCP-PMT is a new type of PMT designed by researchers for the JUNO experiment from the Institute of High Energy Physics of Chinese Academy of Sciences. Fig. 1 shows the structure of the MCP-PMT \cite{lab9}. From the figure, it can be known that there are two groups of MCP stacked in the insulated trestle. Between these two MCP groups is the anode. Signals is led out by a cable in the bracket.

\begin{center}
\includegraphics[width=7cm]{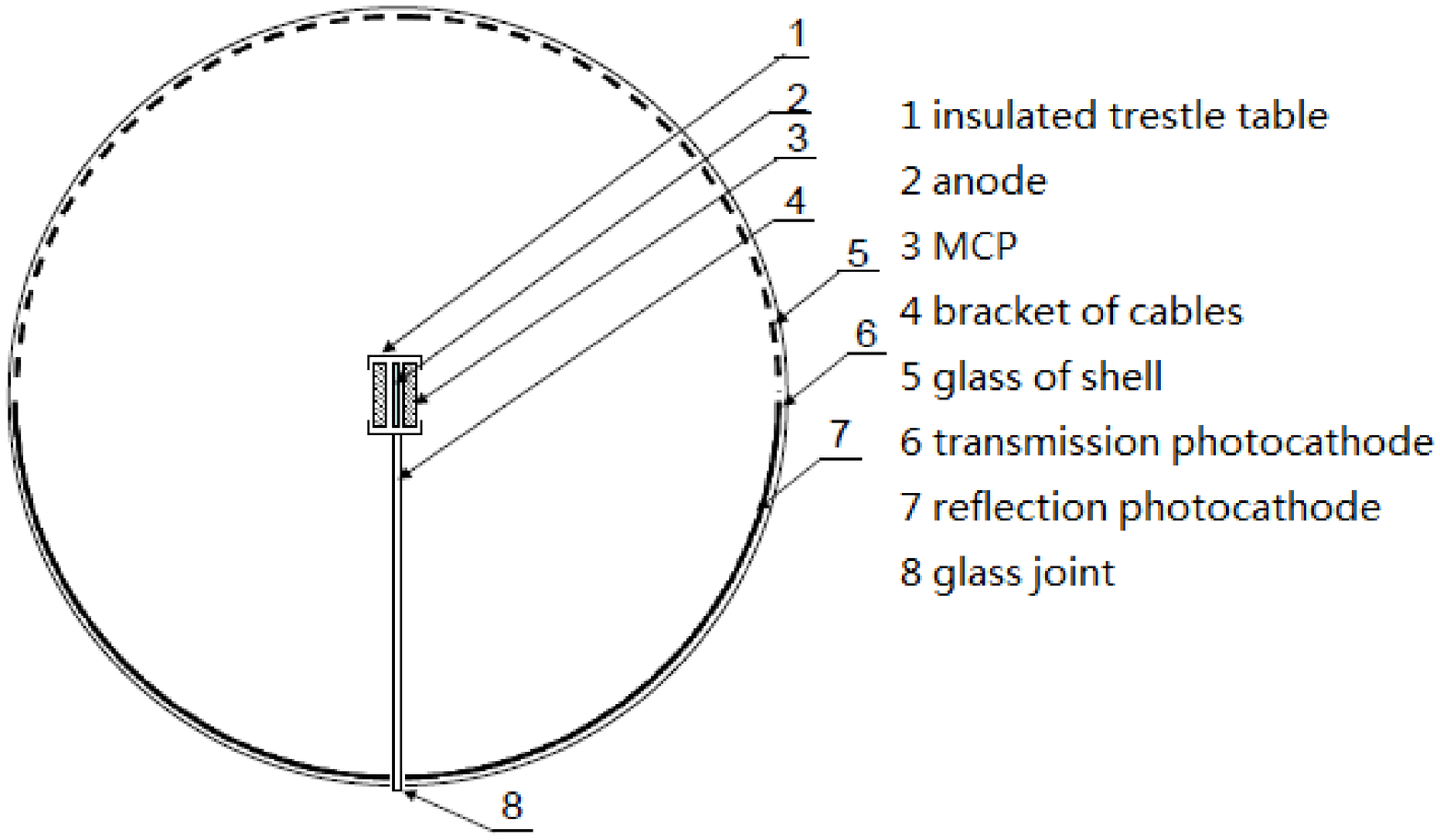}
\figcaption{\label{fig1}  structure of MCP-PMT }
\end{center}

\section{Experiment Setup of MCP-PMT Aging}
\subsection{Reasons and Mechanism of MCP-PMT Aging}
There are two aspects about MCP-PMT aging: the photocathode aging and the MCP aging. The residue gas atoms, and atoms absorbed on the MCP inner hole are ionized by the photoelectron accelerated in the micro channel. The ionized atoms are accelerated by bias voltage and thereby travel to the PMT photocathode, and is known as an ion feedback. The photocathode losses its activity in a process known as photocathode poisoning, when the ionized atoms accumulate on the photocathode itself \cite{lab10}.  Then, electrons are multiplied in the micro channel. When the electrons reach the end of the last micro channel, they form a considerable strong cluster. This cluster may interact with the semiconducting layer of the inner hole of the MCP, and thereby causes damage to it.

As the PMT photocathode is at a short distance from its focusing anode, the life of the proximity-focus PMT is mainly restricted by the photocathode \cite{lab11}. In order to prevent photocathode poisoning, many sophisticated designs have been put forward, such as by using ion blocking film \cite{lab10}, being stacking in ¡®V¡¯ for two MCP \cite{lab12}, being stacking in ¡®Z¡¯ for three MCP , using the vacuum transfer equipment , modifying PMT structure \cite{lab13} or improving the cathode material. Aging of large area MCP-PMT caused by the photocathode poisoning factor is reduced greatly for the large distance between the photocathode and the MCP module. The aging of the photocathode is no longer the main limit of MCP-PMT, while mechanical damage is the main factor of this type of PMT aging.

Many articles studied the aging of proximity-focus PMT \cite{lab11,lab13,lab14}. Relationships have been tested between the quantum efficiency of photocathode and the cumulative anode output charges per unit area, and between the dark count and cumulative anode output charges per unit area. The studies have also conduct aging experiment of different photocathodes and scanned photocathode quantum efficiencies.

The two large-area MCP-PMT prototypes for aging tests were new types of PMT, whereby their multipliers comprised of two groups of MCP, with each MCP group comprising of two MCPs stacking in a ¡®V¡¯ formation. Two groups of MCPs are fixed on the ceramic ring vertically with signals led out by two anode metal wires. The 8 inch prototype of large area MCP-PMT with the tagging number 18\# and 19\# were used for the aging experiment. Depending on the actual condition of the large area MCP-PMT, the relationship between the PMT gain and cumulative anode output charges, and the relationship between the MPE spectrum central address changes and the cumulative anode output charges were measured. MCP resistance was also measured before and after the experiment.

\subsection{Data Acquisition}

Fig.2 shows the electronics system for the aging experiment. The tested PMT was put in a dark box with a laser diode (LD) supplying light. LD¡¯s driver voltage was supplied by a pulse generator. Single-photon state was obtained by modifying the generator parameters and an online labview program was used to modify the driving voltage of LD. Two signals were generated by the generator. One was to supply the LD driving voltage, while the other one, derived from the differential circuit, forms the differential signal. The gate signal obtained by passing the differential signal through the low threshold discriminator, was put into the door of the plug-in interface, and coincided with the signal outputted by the PMT anode to measure its anode output charge. In order maintain the stability of the PMT input light intensity, we placed a XP2020\cite{lab15} opposite the MCP-PMT to monitor the changes of the intensity of LD¡¯s lighting.

\begin{center}
\includegraphics[width=10cm]{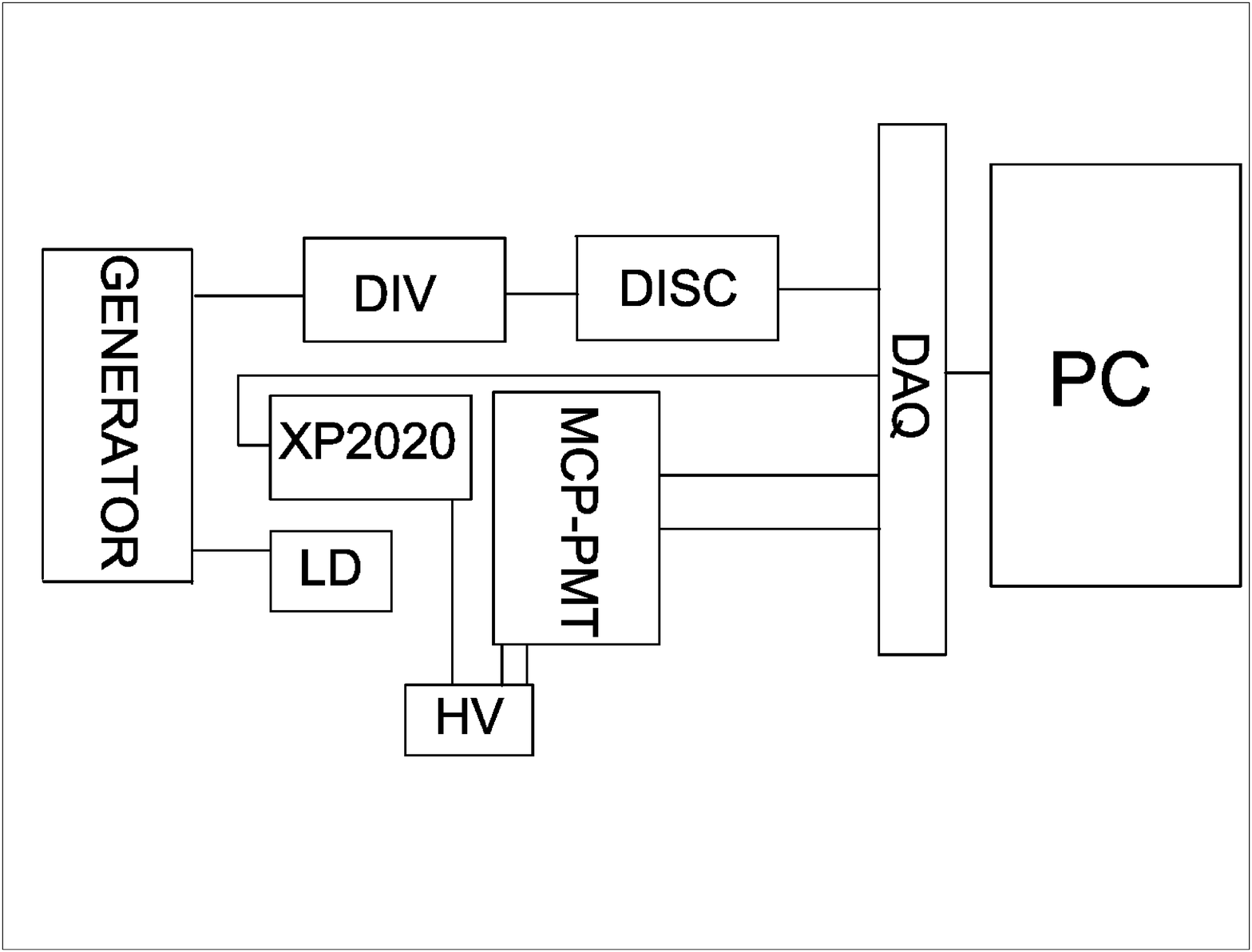}
\figcaption{\label{fig2}  Electronics system of aging experiment }
\end{center}

There are two reasons for using two of the same type of PMTs to conduct aging experiment. Firstly, the experiment data of the two PMTs can be used for cross check, which is more convincing to illustrate PMT aging. Secondly, according to the first PMT aging data, some improvements were made. The high voltage of the second PMT was improved from 1500 V to 1700 V to accelerate aging process, this is equivalent to improve the gain from 1.0$\times10 ^ 7$ to 4$\times  10 ^ 7$, when the cumulative anode output charges were about 4 C.

The PMT detected about 1000 photoelectrons under strong light condition. In the experiment, parameters were tested daily. It¡¯s convenient to switch the light condition between a single photoelectron state and a 1000 photoelectron state. Firstly, MPE spectrum was tested to analyse the whole aging process of the PMT under a strong light condition. The driver¡¯s duty ratio was 0.002\%. Then, SPE spectrum was tested to calculate degradation in the gain of the PMT by modifying the driver¡¯s duty ratio to 0.001\%. Finally, the dark spectrum of the PMT was tested to calibrate the stability of the electronic system by turning off the LD.

\subsection{Stability of Test System}

In the aging period, any small changes of the internal or external environment, such as temperature, light intensity, high voltage and electrons, would have been amplified by the PMT, and would have caused great changes in the MPE spectrum. Thus, the stability of the light source, the PMT itself, and the electron system were monitored in the experiment.

As shown in Fig.2, a XP2020 PMT was put opposite to the tested PMT to monitor the intensity variation of the reflecting light. The XP2020 was running in a low intensity of reflecting light and normal high voltage condition, and hence its aging was very small and could be ignored. Any changes to the XP2020 MPE spectrum could be seen as the changes to the intensity of LD lighting. Fig.3 (a) shows the curve of the MPE spectrum relative to the variation of XP2020. Throughout the experiment, the variation of the MPE mean value of the XP2020 was smaller than 5\%, indicating that the intensity of the LD lighting did not have any observable changes in the experiment process.

The tested PMT was consisted of two MCP groups. In the experiment, the MCP group was exposed to the LD deliberately to study PMT aging process with the other MCP group monitoring internal changes of the PMT, including the internal vacuum and electric field distribution. Fig.3 (b) shows the results of the monitoring PMT¡¯s MPE variation curve. From the Figure, it can be seen that the changes to the monitoring MCP group is less than ¡À5\%, suggesting that internal environment changes of the PMT were negligible.
PMTs and rear electronics were all temperature sensitive. Small changes in the temperature will lead to large changes in the high voltage power supply and PMT gain. The PMTs, high-voltage supply equipment, and electronics were put in the circulation clean room in which the temperature was maintained at 22¡À1¡æ. A CEAN SY1527 high-voltage power supply chassis and A1733 plug-in were used to monitor the changes to the high voltage and current. Fig.3 (c) is the variation in the dark noise spectrum. A horizontal line was found to be a good fit to the data, indicating that the electronic system was stable.

\end{multicols}

\begin{center}
\includegraphics[width=18cm]{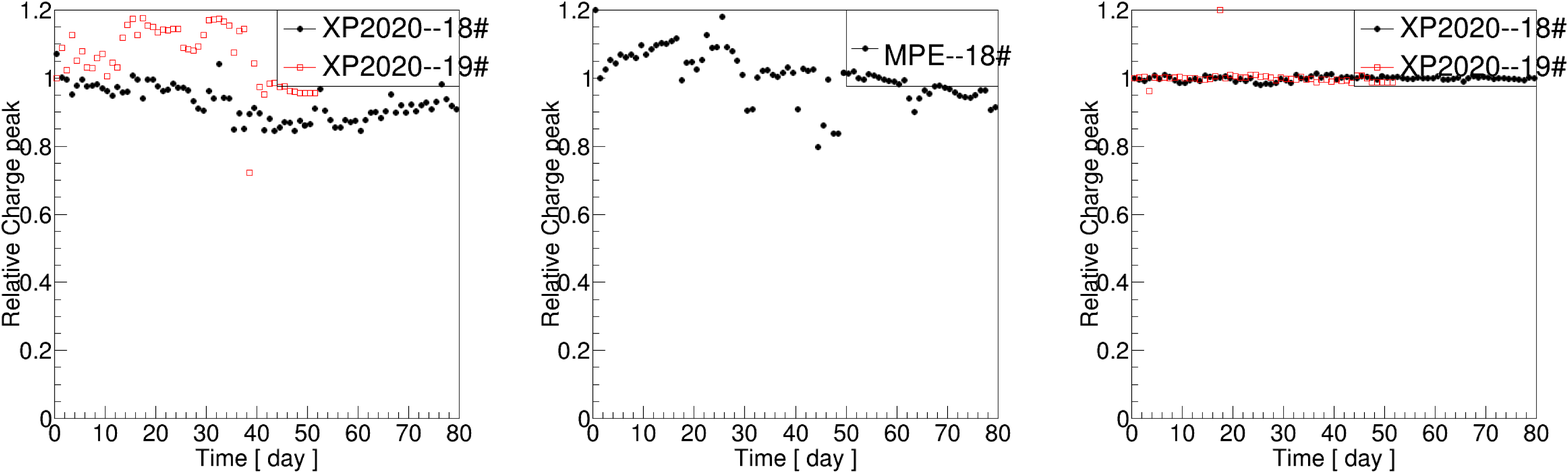}
\figcaption{\label{fig3} Stability of test system}
\end{center}

\begin{multicols}{2}

%

\section{Ageing result and data analysis}
\subsection{Aging of photocathode}

Fig. 4 shows one of the SPE spectrums from the tested MCP-PMT in the experiment. PMT gain was obtained from the difference between the signal peak center and electronic peak center \cite{lab16}.

\begin{center}
\includegraphics[width=7cm]{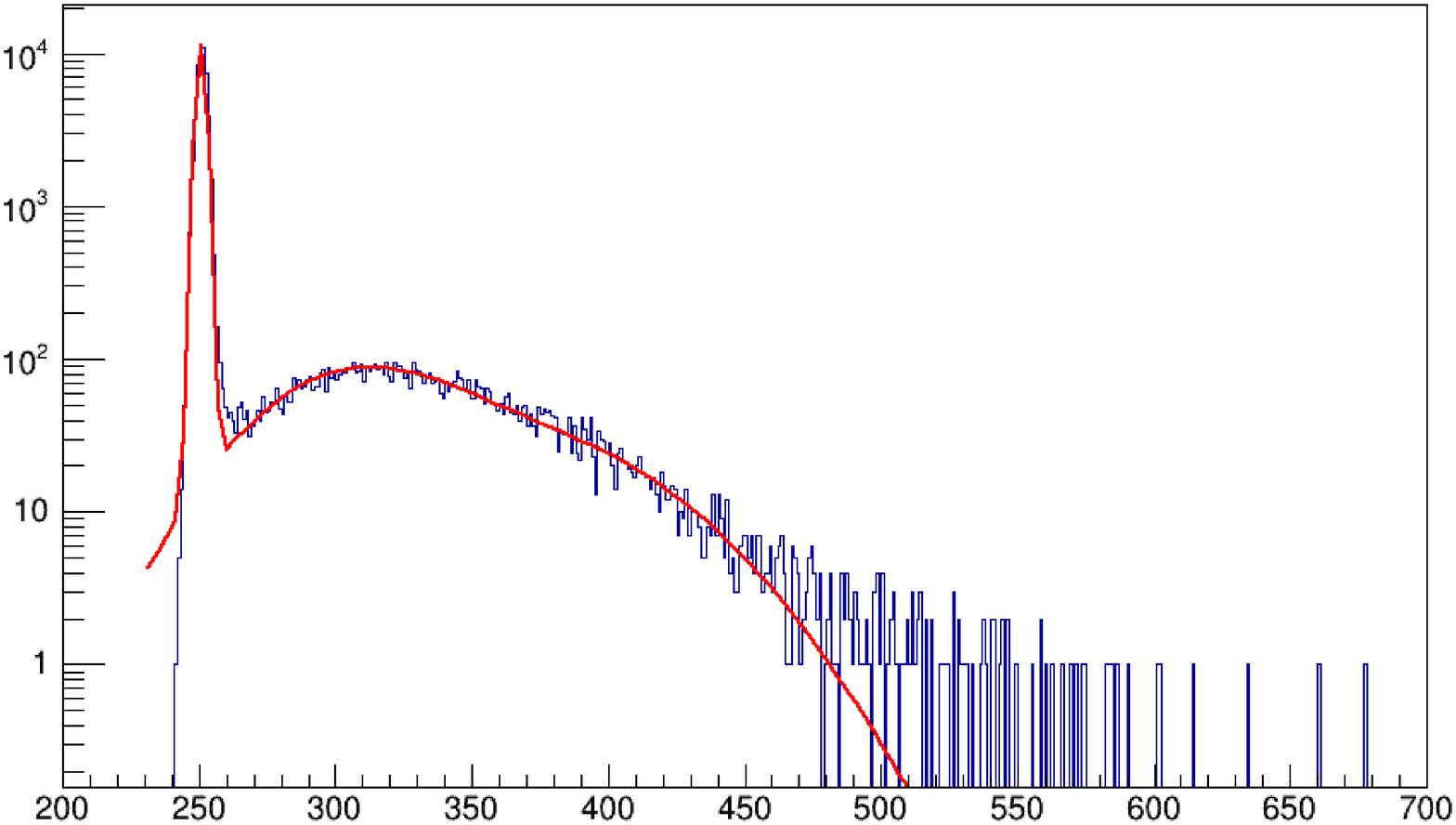}
\figcaption{\label{fig4}  MCP-PMT SPE spectrum }
\end{center}

The number of initial photo electrons, denoted as Npe, was calculated according to the SPE spectrum and MPE spectrum on the first day of test. During the experiment, SPE spectrum and MPE spectrum were tested daily, from which PMT gain, denoted as Gain, the photoelectron loss, denoted as Npl, the measured photoelectrons, denoted as Npx and MPE spectrum center address, denoted as Px can be obtained. The relationship between them can be expressed as:

Npx *Gain = Px.  (1)

Using Npe £¬Npl £¬Npx and Gain, the following can be obtained

(Npx+Npl)*Gain = Npe*Gain.  (2)

From (1) and (2), we get

Px+Npl*Gain = Npe*Gain.  (3)

By rearranging (3), we obtain

Npl=( Npe*Gain- Px)/ Gain.  (4)

Thus, we can get the photocathode relative variation function formula:

F(x) = (Npe - Npl)/ Npe

F(x) = (Npe-( Npe*Gain- Px)/( Gain*Npe)). (5)

Fig. 5 is the PMT photocathode relative variation curve. In the early aging process, the aging speed of photocathode was relatively fast. When the anode cumulative output charge was about 4 C, the change of photocathode was about 10 \%. In the later aging process, there were almost no apparent changes. Up till the cumulative anode output charges were about 18 C at the end of the aging, the relative changes of the photocathode were still around 10 \%.

\begin{center}
\includegraphics[width=7cm]{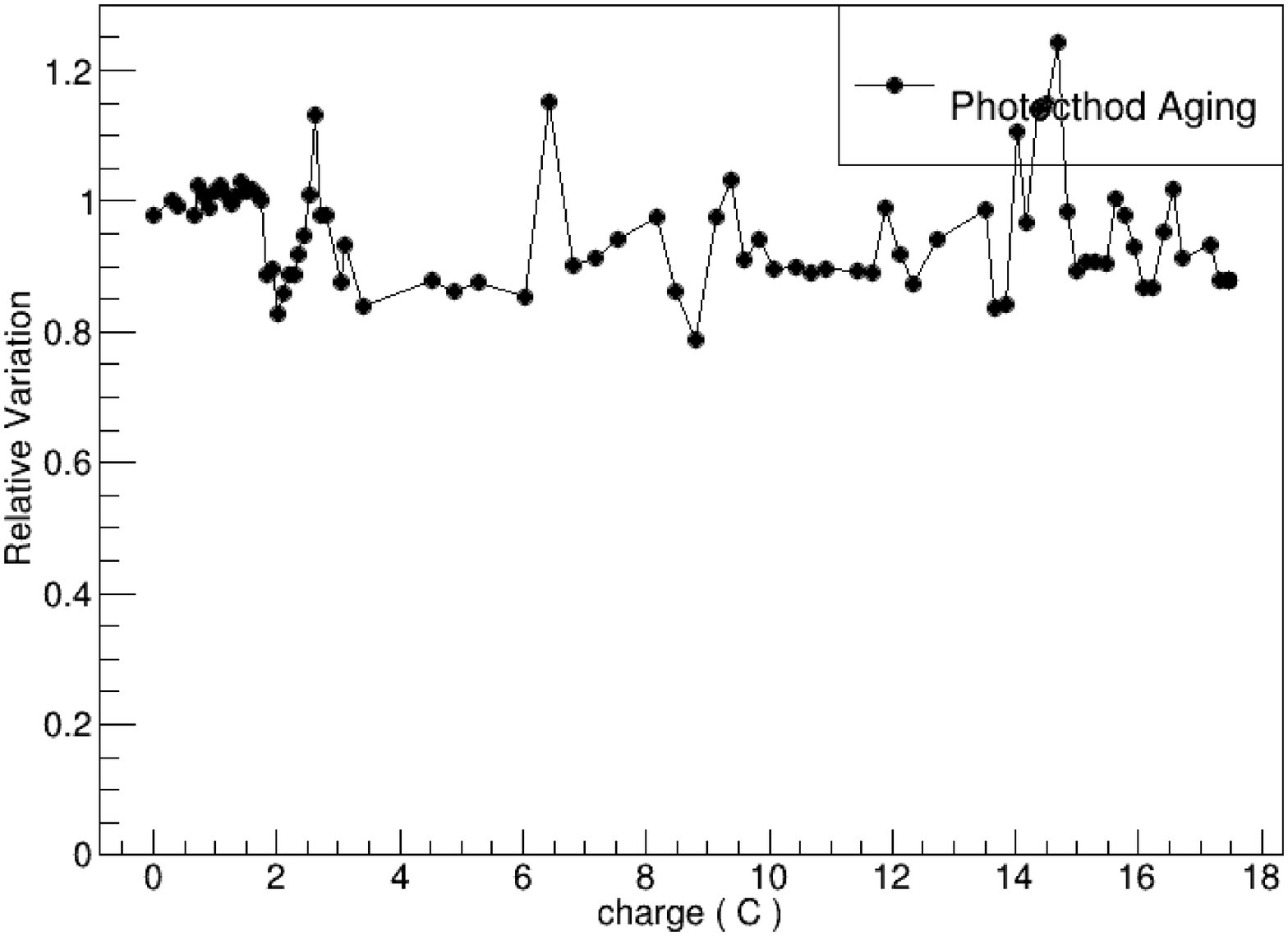}
\figcaption{\label{fig5}  Relative Variation of Photocathode }
\end{center}

\subsection{MCP Body Resistance Variation}

From the aging experiment of the first MCP-PMT, it can be known that aging of it PMT was mainly caused by MCP mechanical damage. MCP mechanical damage might result in changes in the MCP resistance. MCP resistance had influence on HV divided on itself. If the body resistance changed, HV divided on MCP would also had changed, which led to the changes of gain. At the start and end of the experiment, 18\# PMT MCP resistance was tested, and the results are shown in Table 1. From Table 1, it can be observed that the MCP body resistance had no apparent changes after the test. This shows that the gain change was not due to the change in MCP resistance.

\begin{center}
\tabcaption{ \label{tab1}  MCP resistances changes(M¦¸)}
\footnotesize
\begin{tabular*}{80mm}{c@{\extracolsep{\fill}}ccccc}
\toprule
  &\hphantom{0}group 1     &  &\hphantom{0}group 2    &\\
\hline
  & \ MCP1 & \ MCP2 & \ MCP1 & \ MCP2 \\
\ previously  & 145 & 175 & 120 & 150 \\
\ later & 145 & 172 & 120 & 145 \\

\bottomrule
\end{tabular*}
\end{center}

\subsection{Gain Variation of PMT}

The Gain of the MCP-PMT prototype is associated with the lattices of the inner hole of MCP. The aging of MCP can be described by measuring the signal photoelectron output charge. Usually, the secondary electrons are emitted from the MCP lattices when photoelectrons hit the MCP lattices in an electron multiplication process. The lattices will emit little or no secondary electrons from the mechanical damage by the electrons, which will reduce the gain of the MCP at the same operation voltage. Fig. 6 shows the aging process of MCP. The grey part is the conductive electrode at the ends of the channel. The blue section in the middle section is glass base. The yellow inner surface of MCP is a semiconductor component \cite{lab17}. The left part shows the lattices of the inner surface of MCP are undamaged and the right part illustrates the damaged part.

\begin{center}
\includegraphics[width=5cm]{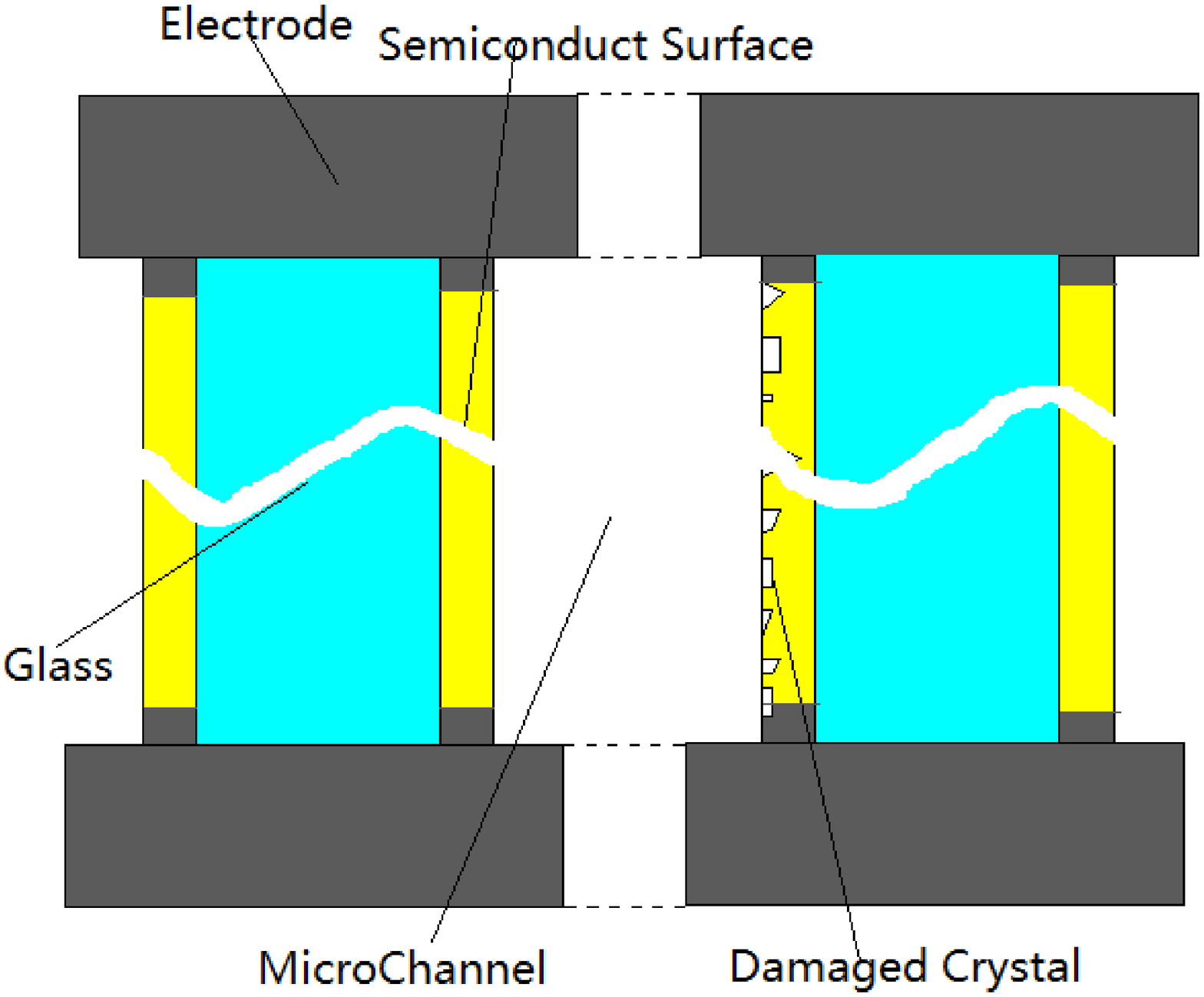}
\figcaption{\label{fig6}  Aging Process of MCP}
\end{center}

The mechanical damage of MCP lattice is a kind of a probabilistic event. The rate of change of the damaged lattice is proportional to product of the number of fine lattices and photoelectron beam¡¯s intensity. That is

¦¤N£½¦Ë1N,   (6)

Where I is a constant. Denoting ¦Ë£±Ias ¦Ë. the equation can be written,

¦¤N£½¦ËN.   (7)

Therefore, MCP fine lattices obey E exponential decay laws, which can be described by an exponential function. PMT¡¯s gain is associated with the rest of the number of fine lattices. In the early period of aging, because the amount of fine lattice was very large, although, there¡¯re many lattices are damaged, but compared to the fine lattice, the damaged lattices can be ignored. So, damaged lattices have less effect on the gain. At the beginning of the aging, the gain of the prototype decreased slowly, and the gain variation (¡÷G1) could be described by the curve as the quadratic function using anode output charge (Q) as variation, as shown in equation (8):

¡÷G1=A1+kQ2,  (8)

Where A1 and K are the constants, and Q is the output charge of MCP.

In the middle and late stages of the aging process, MCP lattices had large changes, and each of lattice damage affected the PMT gain change. Gain variation was in accordance with the lattice E exponential decay, which can be described by E index function using Q as variation, as shown in equation(9):

¡÷G2=A2 * exp (¦Ë Q + A3).   (9)

Where A2, A3 and ¦Ë are constants.

Gain variation function formula can be obtained by integrating the two functions together:

¡÷G= (A1 - KQ2 + A2 $\times$Qn * exp (¦Ë Q + A3)).   (10)

Where A1, A2 and A3 are constants, ¦Ë is the attenuation.  A2$\times$Qn is a regulating factor, such that the quadratic function plays the main role in the early stage of aging, while the exponential function plays the main role in the late stage of aging.

As shown in fig.7, the two curve of the gain variation of prototype 18\# and 19\# can be described with the function as show in (10).

\begin{center}
\includegraphics[width=7cm]{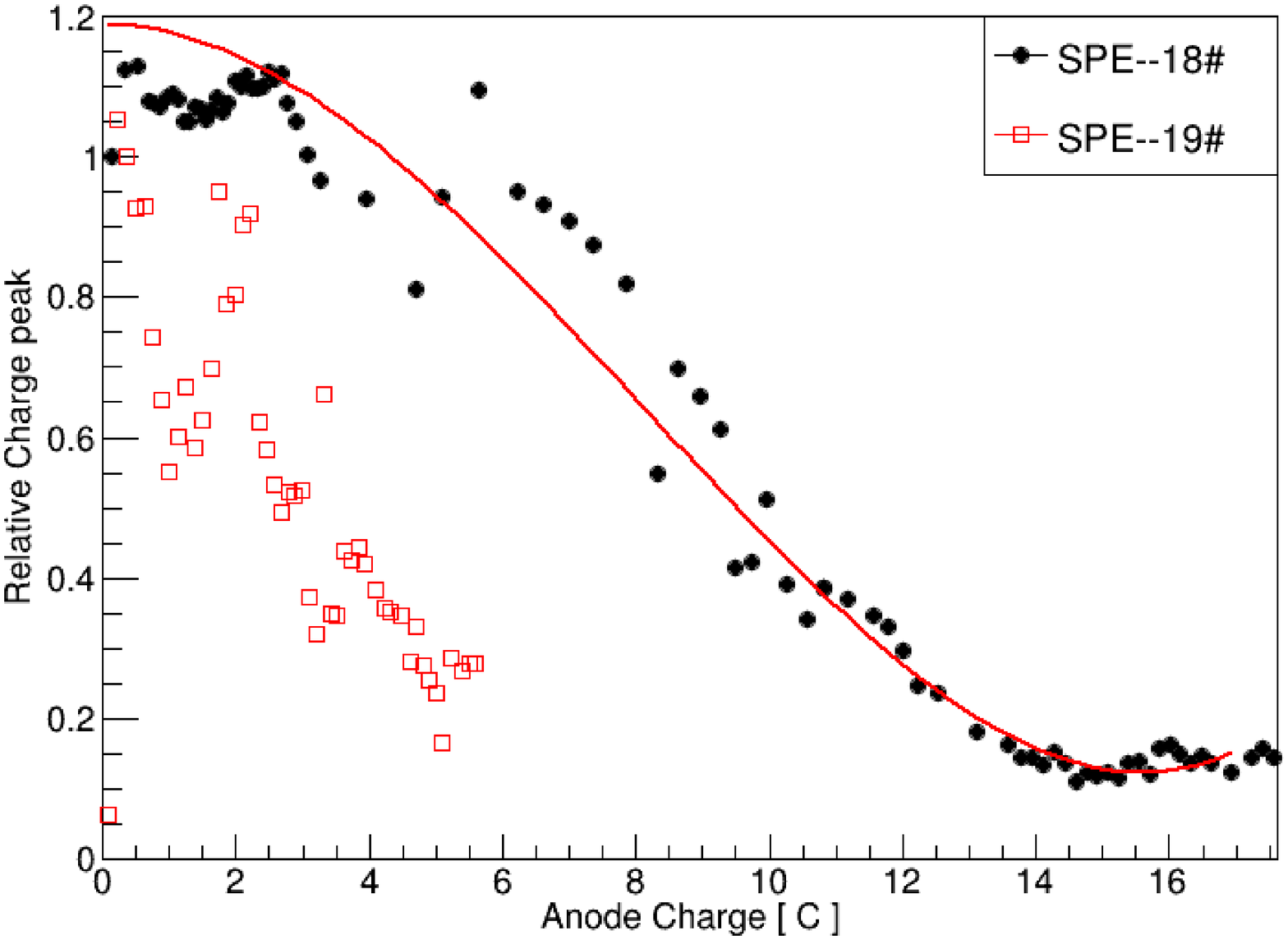}
\figcaption{\label{fig7}  Variation of Gain }
\end{center}

\subsection{ Variation of MPE Spectrum}
The aging of MCP-PMT is a complex process, which includes the mechanical damage of MCP and the degradation of photocathode, and can be described by the relationship between the PMT MPE spectrum mean changes and the cumulative anode output charges. In the early aging period, the change of the photocathode is very small, and the change of the gain is also small. Denoting (¡÷AG1) as the contribution of the change to both stacked, and using Q as a variable to describe in a function of first degree, the relationship of both these variables can be expressed as:

¡÷AG1=A1-KQ.  (11)

In the later aging period, there was no significant alteration of the photocathode of PMT. PMT aging changes were contributed by gain variation, obeying E exponential decay law and can be described by equation (9). Thus, the aging formulation of the PMTs can be written as:

¡÷AG=A1-KQ+A2*exp(-¦ËQ+A3).  (12)

Fig. 8 shows the variation of MPE spectrum. The figure shows that these two PMT changes by 20 \%, when cumulative anode output charge was about 4 C, with the same anode output charge, and the same aging situation in the 18\# and 19\# prototypes. PMT¡¯s aging is observed to comprise of three stages: slow aging in the early stage, rapidly aging in the intermediate stage and and slow in the later stage. When the aging curve changes to about 40 \%, PMT aging slows down again. Comparing Fig. 5, Fig. 8 and Fig. 9, we found that PMT aging is mainly due to the aging of its MCP.

\begin{center}
\includegraphics[width=7cm]{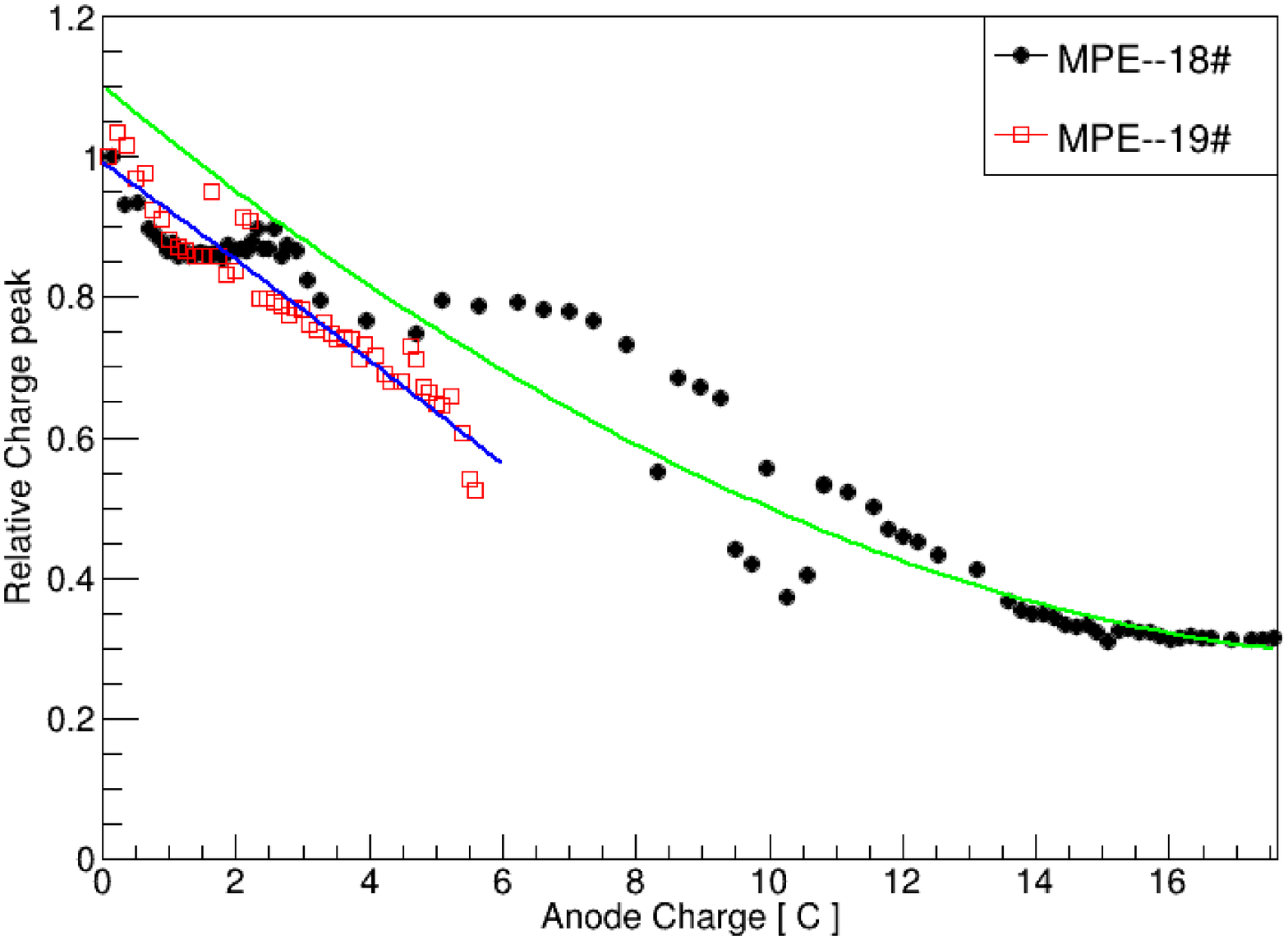}
\figcaption{\label{fig8}  Variation of the MPE spectrum }
\end{center}

\section{ Conclusion}

PMT aging is a complex process,  which is associated with the anode output charge, and is associated with factors, including the MCP, photocathode and the internal vacuum degree. It is known that the aging of large area MCP-PMT is mainly related to the mechanical damage of the last piece of MCP, which can be formulated as a function of the output charge.
If JUNO experiment requires PMT to run more than 10 years with an acceptable change of within 50 \% corresponding to a position on the aging curve at a cumulative anode output charge of about 10 C. Anode signal is mainly composed of background radiation, cosmic rays, and the noise caused by itself. JUNO neutrino group has simulated that signals caused by radiation from the glass of the detector itself, water and rock, cosmic rays, to a total of 2.06 kHz, which when multiplied by the detection efficiency, found that the signal caused by the said sources were at an effective rate of 0.206 kHz. Consequently, the PMT intrinsic noise frequency can be calculated and should to be less than 20 kHz at a gain of 1.0$\times10^7$.

\section{Acknowledgments}

The project was supported by the National Natural Science Foundation of China (Grant No. 10775181 \& No.11475209), and the Strategic Priority Research Program of the Chinese Academy of Sciences (Grant No. XDA10010200 \& No.XDA10010400).

\end{multicols}
\vspace{-1mm}
\centerline{\rule{180mm}{0.1pt}}
\vspace{2mm}

\begin{multicols}{2}

\end{multicols}

\clearpage

\end{document}